\documentclass[useAMS,usenatbib]{mn2e}

\def\aj{AJ}%
\def\apj{ApJ}%
\def\apjl{ApJ}%
\def\apjs{ApJS}%
\def\aap{A\&A}%
\def\aaps{A\&AS}%
\def\mnras{MNRAS}%
\def\pasp{PASP}%
\usepackage{times}
\usepackage[dvips]{graphicx}
\title[Properties of Near-Infrared Selected AGNs]
{Properties of Near-Infrared Selected AGN Candidates with 2MASS/ROSAT Catalogues}
\author[S. Kouzuma and H. Yamaoka]{Shinjirou Kouzuma$^{1}
$\thanks{E-mail:kouzuma@phys.kyushu-u.ac.jp} and Hitoshi Yamaoka$^{1}$\\
$^{1}$Graduate School of Sciences, Kyushu University, Fukuoka 812-8581, Japan}
\begin{document}

\date{Accepted  Received ; in original form }

\pagerange{\pageref{firstpage}--\pageref{lastpage}} \pubyear{2010}

\maketitle

\label{firstpage}

\begin{abstract}
We report on the near-infrared selected AGN candidates extracted
 from 2MASS/ROSAT catalogues and discuss their properties. 
First, near-infrared counterparts of a X-ray source in ROSAT catalogues 
(namely, Bright Source Catalogue (BSC) and Faint Source Catalogue (FSC))
 were extracted by positional cross-identification of $\leq 30''$. 
Because these counterparts would contain many mis-identifications, 
we further imposed near-infrared colour selection criteria and 
extracted reliable AGN candidates (BSC: 5,273, FSC: 10,071). 
Of 5,273 (10,071) candidates in the BSC (FSC), 2,053 (1,008) are known AGNs. 
Near-infrared and X-ray properties of candidates show similar properties with known AGNs and
 are consistent with previous studies. 
We also searched for counterparts in other wavelengths (that is, optical, near-infrared, and radio), and 
investigated properties in multiwavelength. 
No significant difference between known AGNs and unclassified sources could be seen. 
However, some unclassified sources in the FSC showed slightly different properties compared with known AGNs. 
Consequently, it is highly probable that we could extract reliable AGN candidates
 though candidates in the FSC might be spurious. 
\end{abstract}


%%%%%%%%%%%%%%%%%%%%%%%%%%%%%%%%%%%%%%%%
\section{Introduction}
Active galactic nuclei (AGNs) are luminous in wide wavelength 
because of the vast amount of energy produced by accretion onto a central supermassive blackhole. 
Accordingly, they are observable at nearly entire wavelength despite locating at great distances, and
 have been searched in various wavelengths: 
e.g., optical \citep{Schneider2007-AJ,Veron2006-AA}, infrared \citep{Low1988-ApJ,Low1989-ApJ},
or radio \citep{Frayer2004-AJ}. 
X-ray emission is especially a characteristic property of AGNs. 
The vast majority of X-ray sources are AGNs
 and all classes of AGNs appear in X-ray surveys. 
There are many studies that collect AGN samples using X-ray data
 \citep[e.g., ][]{Kim1999-ApJ,Watanabe2004-ApJ,Polletta2007-ApJ}. 
Therefore, whether an object is a X-ray source can be a criterion to select AGNs. 

Colour selection is a powerful technique in extracting AGN candidates. 
A classical method is known as the $UV$-excess \citep[UVX; ][]{Sandage1965-ApJ,Schmidt1983-ApJ,Boyle1990-MNRAS}, 
which extract bluer quasars. 
\citet{Richards2002-AJ} selected quasars via their nonstellar colours using SDSS photometry. 
Other selections are such as red quasar survey
 using optical and near-infrared combined colours \citep{Glikman2007-ApJ}, or 
mid-infrared selected AGNs using Spitzer data \citep{Lacy2004-ApJS,Stern2005-ApJ}.  

Colour selection using near-infrared photometry has also performed by some previous studies. 
\citet{Cutri2001-ASPC,Cutri2002-ASPC} extracted obscured AGN candidates
 using the colour selection of $(J-K_\textnormal{\tiny S})>2.0$. 
The $KX$ method using the excess in K-band, proposed by \citet{Warren2000-MNRAS}, was used for extracting quasars
 \citep{Jurek2008-MNRAS,Maddox2008-MNRAS,Smail2008-MNRAS,Nakos2009-AA}. 
However, these extracted only peculiar AGNs (in the former case) or
 extracted AGNs using combined with optical photometry (in the latter case). 
Because the optical light suffers more extinctions than the infrared light and an AGN is surrounded by a dust torus, 
a selection using optical and near-infrared combined colours may miss AGNs (especially obscured AGNs), 
 because of lack of optical detection. 
However, \citet{Kouzuma2010-AA} proposed colour selection criteria
 to extract AGNs using only near-infrared colours. 
They demonstrated by both observed and simulated colours that
 AGNs are differentiated from several types of objects
 in a $(H-K_\textnormal{\tiny S})$-$(J-H)$ colour-colour diagram (CCD). 
This enables us to extract AGN candidates using only near-infrared photometry. 

In this paper, we first extract bright sources in both near-infrared and X-ray
 by a cross-identification between the Two Micron All Sky Survey (2MASS) and ROSAT all-sky survey catalogues, and 
select AGN candidates on the basis of the near-infrared colour selection criteria proposed by \citet{Kouzuma2010-AA}. 
In addition, we investigate properties of candidates using not only near-infrared and X-ray data
 but also photometric data at other wavelengths derived by cross-identifications with some catalogues. 
In Section \ref{DATA}, we introduce the 2MASS and ROSAT. 
In Section \ref{EXTRACTION}, we describe the method to extract AGN candidates (including the criteria of both cross-identification and colour selection) and 
the results. 
In Section \ref{PROPERTIES}, properties of AGN candidates are investigated by photometric data at 
near-infrared, X-ray, and other wavelengths.

%%%%%%%%%%%%%%%%%%%%%%%%%%%%%%%%%%%%%%%
\section{Data}\label{DATA}
%%%%%%%%%%%%%%%%%%%%%%%%%%%%%
\subsection{2MASS}
The 2MASS\footnote{2MASS web site (http://www.ipac.caltech.edu/2mass/)} \citep{Skrutskie2006-AJ}
is a project that observed 99.998\% of the whole sky 
at J (1.25 $\mu$m), H (1.65 $\mu$m), Ks (2.16 $\mu$m) bands, 
at Mt. Hopkins, AZ (in the Northern Hemisphere) and at CTIO, Chile (in the Southern Hemisphere)
between 1997 June and 2001 February. 
The instruments are both highly automated 1.3-m telescopes equipped with three-channel cameras, 
each channel consisting of a 256 $\times$ 256 array of HgCdTe detectors. 
The 2MASS obtained 4,121,439 FITS images (pixel size $\sim2''_{\cdot}0$) with 7.8 s of integration time.
The limiting magnitudes (Signal-to-Noise Ratio (S/N)$>$10) are 
15.8 (J), 15.1 (H), 14.3 (K$_\textnormal{\tiny S}$) mag at each band.
The Point Source Catalogue (PSC) was produced using these images and catalogued 470,992,970 sources.
In the 2MASS web site, the images and the PSC are open to the public and are easily available.

%%%%%%%%%%%%%%%%%%%%%%%%%%%%%
\subsection{ROSAT}
The ROSAT is an X-ray astronomical satellite launched in 1990.
During six months soon after the launching, 
all-sky imaging survey in the energy range of $0.1-2.4$keV was performed with a X-ray reflector 
equipped with Position Sensitive Proportional Counter (PSPC). 
A limiting PSPC count-rate is $0.05$ cts s$^{-1}$. 
This is the first imaging survey in the X-ray wavelength.
The survey has yielded Bright Source Catalogue (BSC) and Faint Source Catalogue (FSC), 
and they contain 18,806 and 105,926 sources, respectively, with positional error of $\la 30''$.
Both catalogues have already been open to the public and available on the web.
In this paper, we use both BSC and FSC.

%%%%%%%%%%%%%%%%%%%%%%%%%%%%%%%%%%%%%%%
\section{Extraction of AGN candidates}\label{EXTRACTION}
%%%%%%%%%%%%%%%%%%%%%%%%%%%%%
\subsection{Cross-identification}
We cross-identified ROSAT catalogues with the 2MASS PSC to extract a near-infrared counterpart for a X-ray source. 
We set a positional criterion of $\leq 30''$, that is, 
we treat as a near-infrared counterpart when a 2MASS source is within $30''$ of a ROSAT source. 
As a result of the cross-identification, we extracted 46,205 (217,033) sources. 
Even though several 2MASS sources may be located within $30''$ of a ROSAT source, 
we treated them as counterparts for the X-ray source. 
Accordingly, in the positional criterion of $\leq 30''$,
 many objects are miss-identified due to the difference of angular resolutions. 
Therefore, we further imposed colour selection criteria to extract reliable counterparts. 

%------------------
\subsection{Colour Selection}
The difference between positional accuracies causes mis-identification on a cross-identification. 
Therefore, another criterion is required for accurately extracting a counterpart. 
\citet{Bessell1988-PASP} presented stellar locus 
in a near-infrared ($H-K_\textnormal{\tiny S}$)-($J-H$) CCD. 
Because most 2MASS sources are galactic normal stars, 
mis-identified sources should be located around the stellar locus in the near-infrared CCD. 
\citet{Kouzuma2010-AA} investigated near-infrared colours of quasars/AGNs and 
demonstrated that the locus of quasars/AGNs are differentiated from the stellar locus in the near-infrared CCD. 
They also proposed near-infrared colour selection criteria for extracting AGNs: 
\begin{equation}
(J-H) \leq 1.70(H-K_\textnormal{\tiny S})-0.118
\end{equation}
\begin{equation}
(J-H) \leq 0.61(H-K_\textnormal{\tiny S})+0.50
\end{equation}
We adopted these criteria for extracting reliable AGN candidates 
from the cross-identified sources. 
Before the near-infrared colour selection, 
we extracted the sources having photometric quality flags in the 2MASS PSC superior to B 
(corresponding to $S/N>7$). 
Samples were reduced to 27,058 (121,767) sources. 
We subsequently extracted AGN candidates on the basis of the near-infrared colour selection criteria. 

Because most sources with $K_\textnormal{\tiny S}<10$ are probably normal stars, 
we further ruled out such sources (i.e., extracted only sources with $K_\textnormal{\tiny S} \geq 10$).

%------------------
\subsection{Extracted Candidates}

Finally, we have extracted 5,273 (10,701) sources as AGN candidates in the BSC (FSC). 
To confirm that candidates have been already known,
 we checked the following AGN catalogues: 
\citet{Veron2006-AA,Schneider2007-AJ,Anderson2007-AJ,Brinkmann2000-AA,Wei1999-AAS,Yuan1998-AA}. 
%\citet{Moran1996-ApJS,Brinkmann2000-AA,Bauer2000-ApJS,Grupe2004-AJ,
%VeronCetty2004-AA,Veron2006-AA,Schneider2007-AJ,Mullis2004-ApJ}. 

Table \ref{AGN-catalogues} shows the number of identified sources in each catalogue. 
Of 5,273 (10,071) candidates, 2,053 (1,008) sources have been already known as AGNs. 
It is highly probable that remaining 3,220 (9,693) sources are unknown AGNs. 
Below, we consider the reliability of AGN candidates by investigating properties of them. 

\begin{table*}
\begin{center}
\caption{The number of known AGNs in our AGN candidates. 
 \label{AGN-catalogues}}
\begin{tabular}{lrr}
\hline \hline
AGN catalogue & BSC & FSC \\ \hline
Quasars and Active Galactic Nuclei (12th Ed.) \citep{Veron2006-AA} & 1,900 & 871 \\
SDSS-DR5 quasar catalog \citep{Schneider2007-AJ} & 493 & 343 \\
AGN from RASS and SDSS DR5 \citep{Anderson2007-AJ} & 776 & 537 \\
ROSAT-FIRST AGN correlation \citep{Brinkmann2000-AA} & 171 & 108 \\
RASS AGN sample \citep{Wei1999-AAS} & 110 & 0 \\
ROSAT detected quasars II \citep{Yuan1998-AA} & 104 & 32 \\
total known sources & 2,053 & 1,008 \\ \hline
\end{tabular}
\end{center}
\end{table*}

%%%%%%%%%%%%%%%%%%%%%%%%%%%%%%%%%%%%%%%
\section{Properties of AGN candidates}\label{PROPERTIES}
%%%%%%%%%%%%%%%%%%%%%%%%%%%%%
\subsection{Near-Infrared and X-ray Properties}

We investigate near-infrared and X-ray properties of the AGN candidates. 
The zero-magnitudes of near-infrared fluxes at three bands are based on \citet{Cohen2003-AJ}. 
The ROSAT PSPC count rate (0.1 -- 2.4 keV) of each source was converted to a X-ray flux 
by assuming that the spectrum is a power-law with an index of $N=-2$. 

%%%%%%%%%%%%%%%%%%%%%%%%%%%%%
\subsubsection{Photometric Properties}

\begin{figure}
	\begin{center}
		\resizebox{40mm}{!}{\includegraphics[clip]{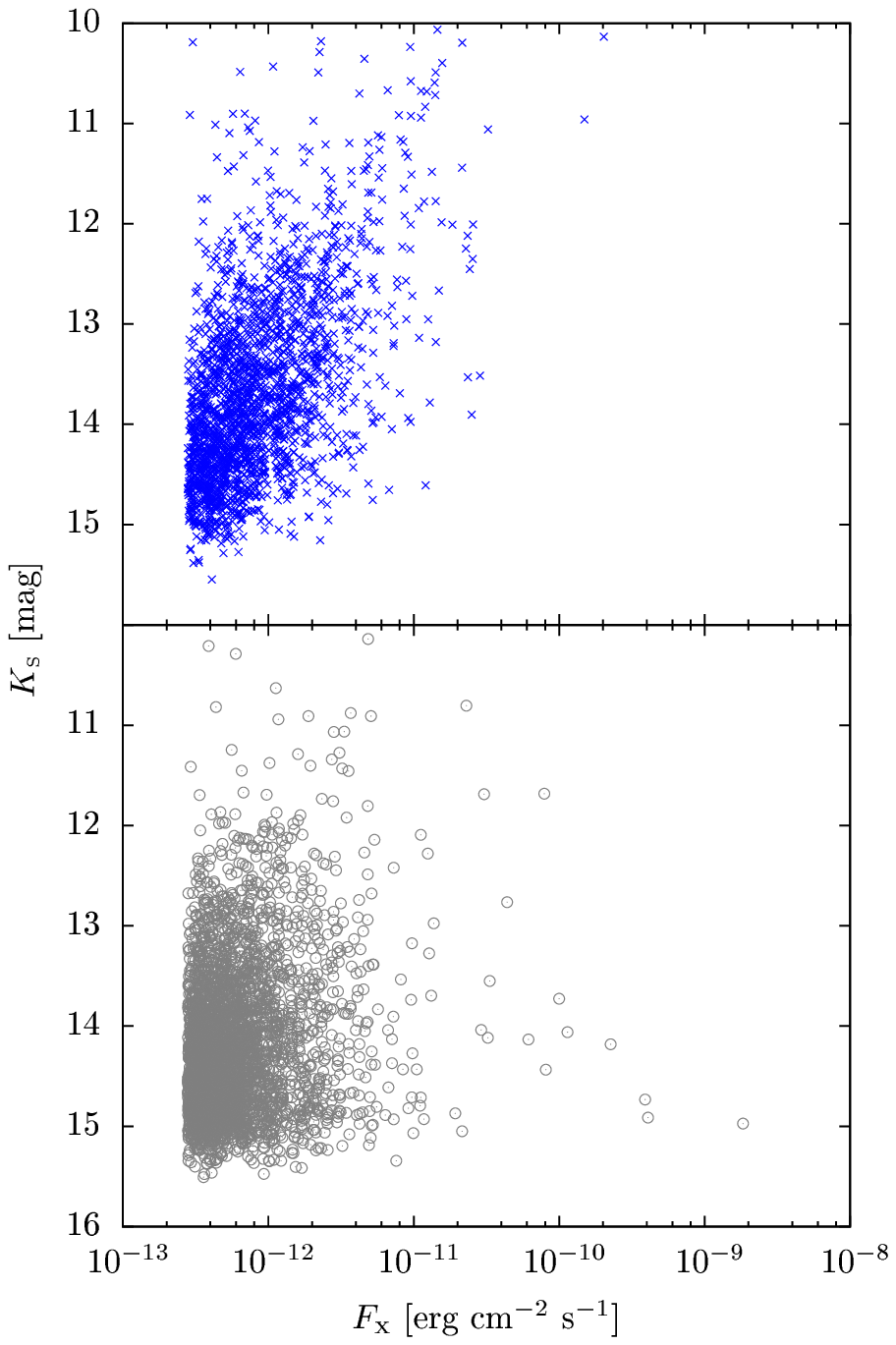}}
		\resizebox{40mm}{!}{\includegraphics[clip]{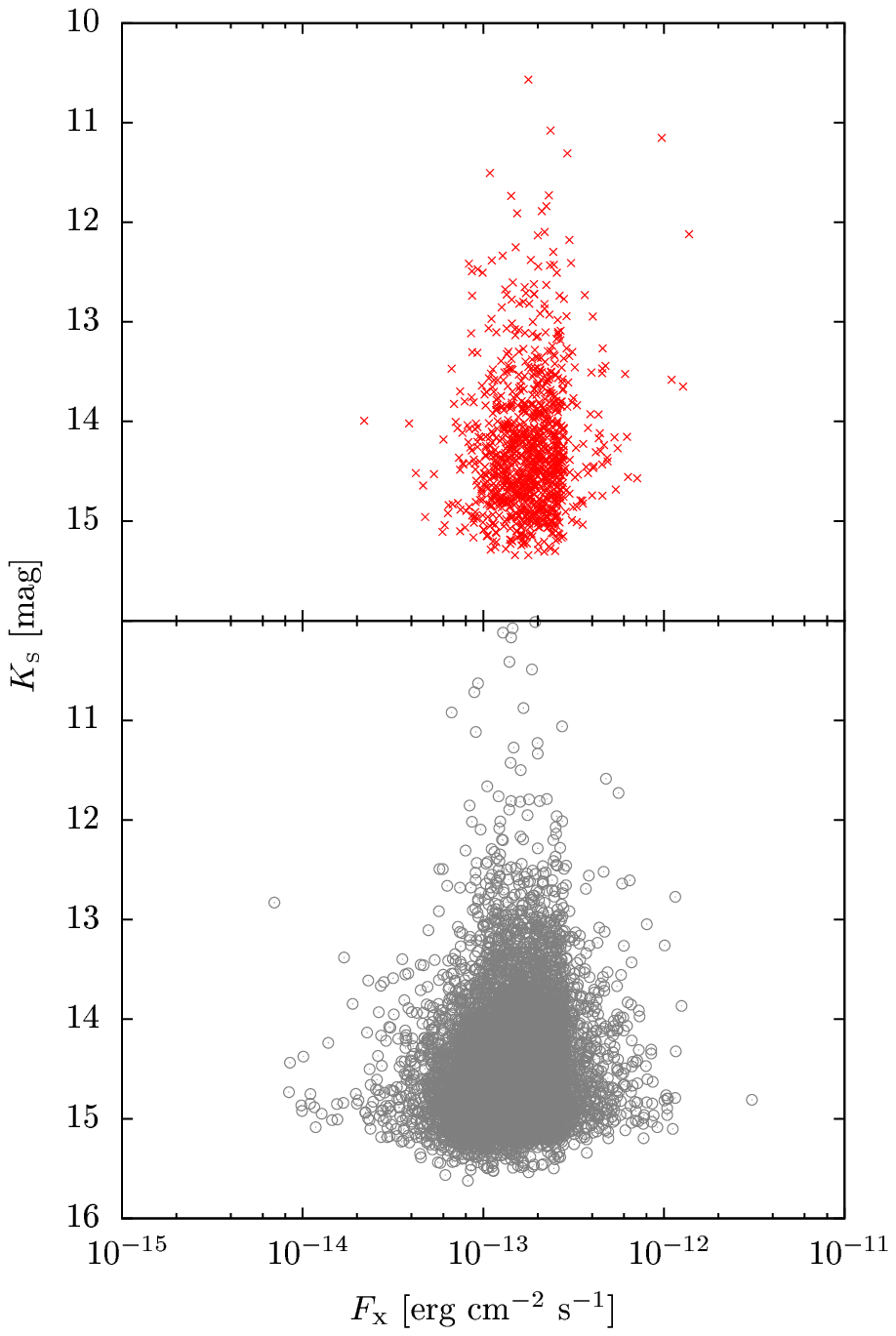}}
	\end{center}
		\caption{X-ray flux versus K$_\textnormal{\tiny S}$-band magnitude. 
					The left panel is the distributions of candidates in the BSC and 
					the right panel is those in the FSC. 
					The symbols represent known AGNs in the BSC (blue crosses), 	
					known AGNs in the FSC (red crosses), and unclassified sources in both BSC and FSC (gray circles). 
  \label{Fx-K}}
\end{figure}

Figure \ref{Fx-K} shows the $F_\textnormal{\tiny X}$-$K_\textnormal{\tiny S}$ diagrams. 
It is known, by previous studies, that there is a correlation between 
optical magnitude and X-ray flux. 
For known AGNs in the BSC, there appears to be a correlation between $K_\textnormal{\tiny S}$ and X-ray flux. 
The distribution of unclassified sources is similar to that of known AGNs, 
although a correlation of unclassified sources seems to be weak. 
Because FSC sources are mainly gathered sources fainter than the count rate of $\sim0.05$ counts s$^{-1}$, 
there is a small number of sources in the range of $F_\textnormal{\tiny X}>2.8 \times 10^{-13}$ 
erg cm$^{-2}$ s$^{-1}$ (corresponding to 0.05 counts s$^{-1}$).

\begin{figure}
	\begin{center}
		\resizebox{40mm}{!}{\includegraphics[clip]{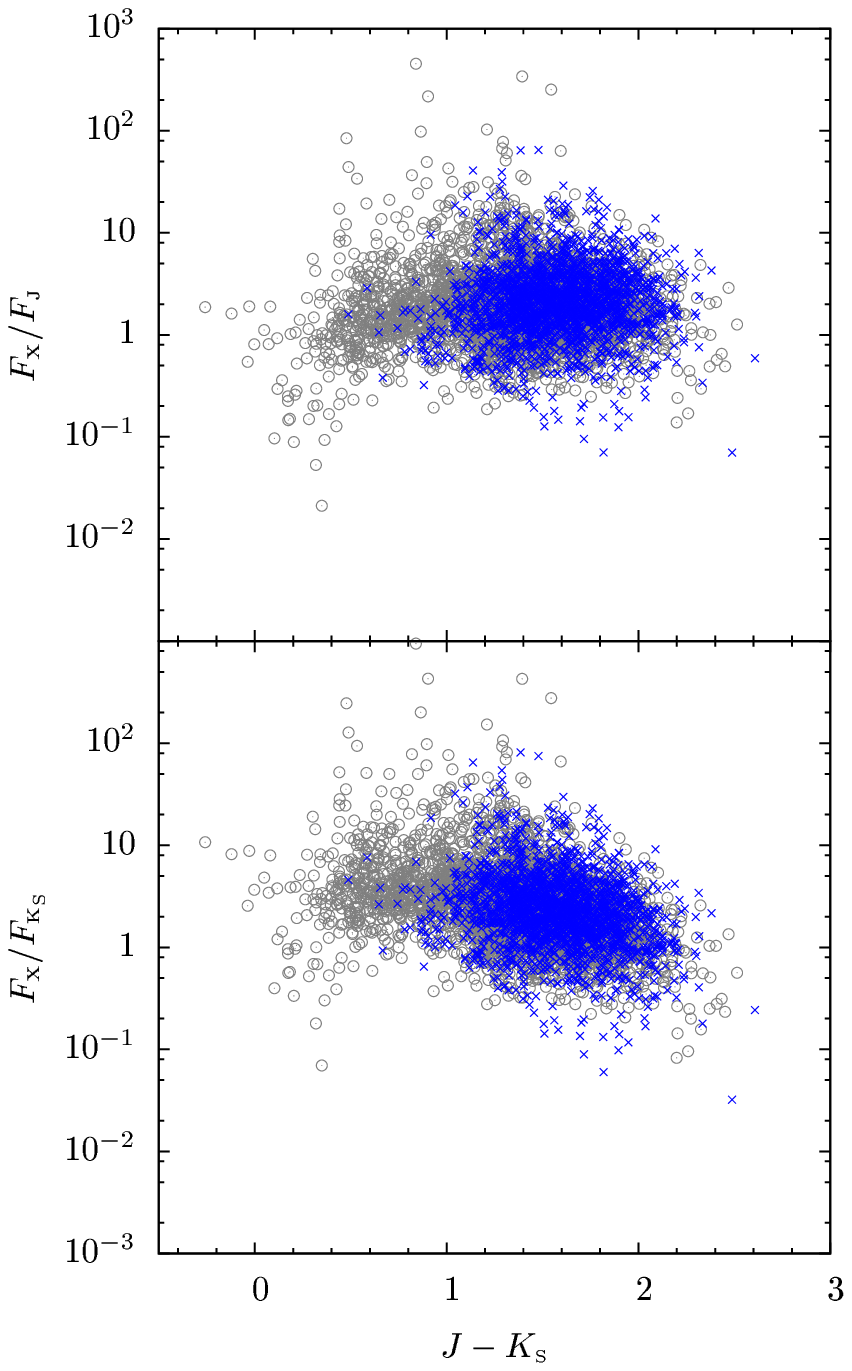}}
		\resizebox{40mm}{!}{\includegraphics[clip]{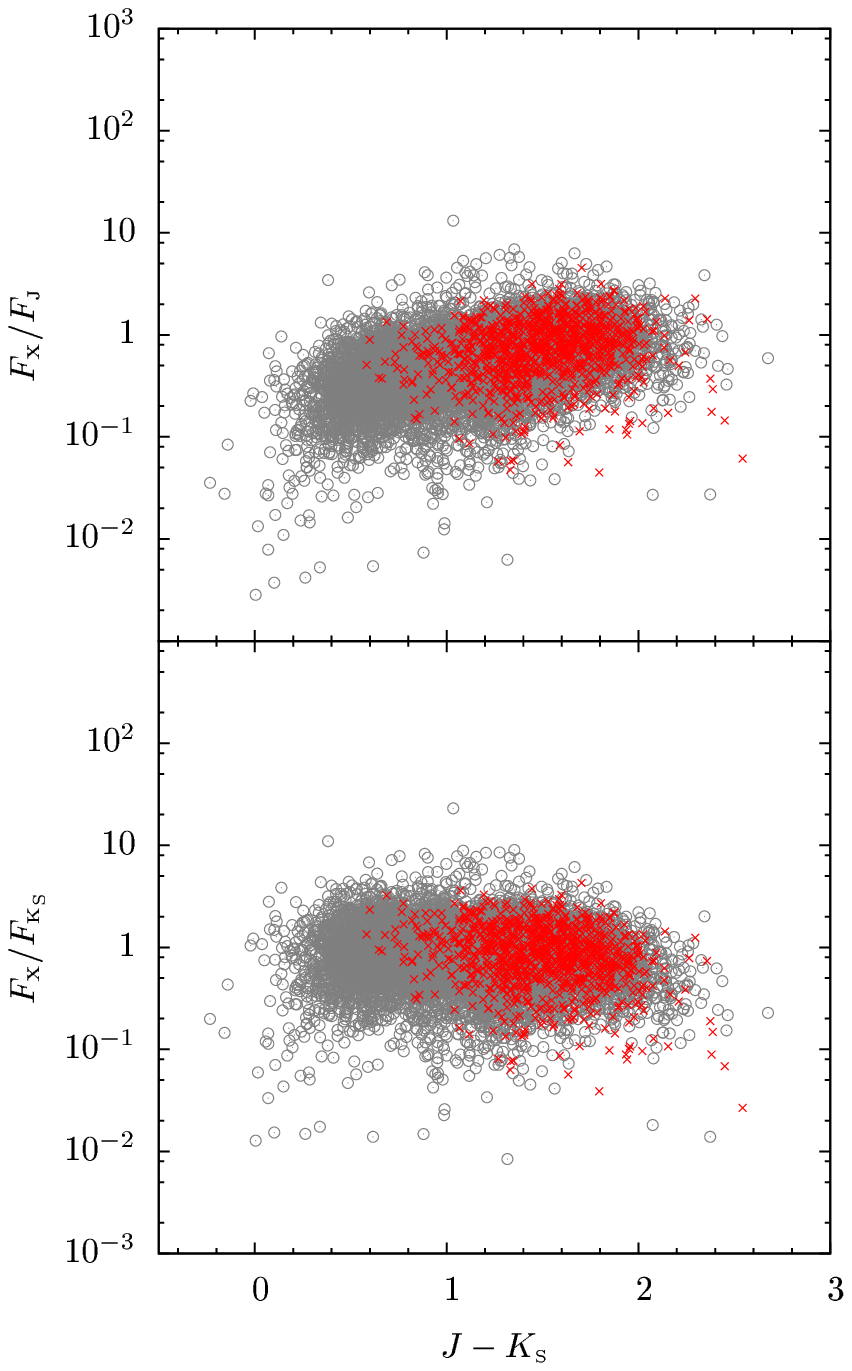}}
	\end{center}
		\caption{Near-infrared colour versus X-ray to J-band flux ratio diagram. 
					The symbols are the same as in Figure \ref{Fx-K}. 
  \label{JK-XJ-CCD}}
\end{figure}

Figure \ref{JK-XJ-CCD} shows the ($J-K_\textnormal{\tiny S}$)-
($F_\textnormal{\tiny X}/F_\textnormal{\tiny J}$) diagrams. 
As to candidates in the BSC, $F_\textnormal{\tiny X}/F_\textnormal{\tiny J}$ is nearly constant 
though $F_\textnormal{\tiny X}/F_\textnormal{\tiny K$_\textnormal{\tiny S}$}$ 
decreases with $(J-K_\textnormal{\tiny S})$ colour. 
On the other hand, as to candidates in the FSC,
  $F_\textnormal{\tiny X}/F_\textnormal{\tiny J}$ is directly proportional to $(J-K_\textnormal{\tiny S})$ colour
 and $F_\textnormal{\tiny X}/F_\textnormal{\tiny K$_\textnormal{\tiny S}$}$
 decreases with  $(J-K_\textnormal{\tiny S})$ colour compared with
 $F_\textnormal{\tiny X}/F_\textnormal{\tiny J}$. 
A similarity between candidates in the BSC and the FSC is that
 $F_\textnormal{\tiny X}/F_\textnormal{\tiny K$_\textnormal{\tiny S}$}$ 
decreases with $(J-K_\textnormal{\tiny S})$ colour.  This is probably due to extinctions 
because the light at the J-band suffers extinction than the light at the K$_\textnormal{\tiny S}$ band 
(i.e., $F_\textnormal{\tiny J}$ should be absorbed than $F_\textnormal{\tiny K$_\textnormal{\tiny S}$}$). 

\citet{Haakonsen2009-ApJS} reported loci of various types of objects in a
 ($J-K_\textnormal{\tiny S}$)-($F_\textnormal{\tiny X}/F_\textnormal{\tiny J}$) diagram
 using both 2MASS photometry and X-ray flux in the ROSAT BSC. 
Their diagram shows that the AGN locus is well separated from those of other types of objects
 such as normal stars,
 where almost all of the AGNs have $(J-K_\textnormal{\tiny S})>0.6$ and 
$F_\textnormal{\tiny X}/F_\textnormal{\tiny J}>3 \times 10^{-2}$. 
The locus of our candidates is consistent with  their AGN locus. 

Some unclassified sources in the BSC or the FSC have $(J-K_\textnormal{\tiny S})<1.0$, 
where few known AGNs are distributed. 
These sources might be spurious AGNs or 
other kinds of AGNs (e.g., AGNs having extremely small extinction and low luminosity) 
that have not been found by previous surveys. 

The AGNs with $(J-K_\textnormal{\tiny S})>2.0$ are defined as red AGNs \citep{Cutri2001-ASPC}. 
It is believed that many red AGNs are found at $z\la 0.5$ \citep{Cutri2002-ASPC}. 
In our sample, there are 234 (192) red AGNs in the BSC (FSC) and
 it is highly probable that they are AGNs at $z \la 0.5$. 

%------------------
\subsubsection{Hardness Ratio}

\begin{figure}
	\begin{center}
		\resizebox{40mm}{!}{\includegraphics[clip]{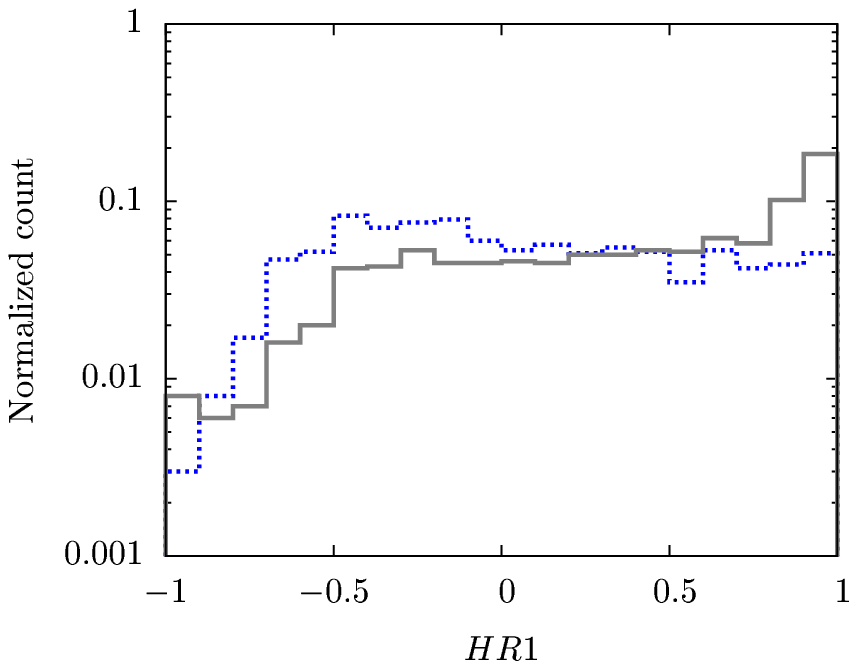}}
		\resizebox{40mm}{!}{\includegraphics[clip]{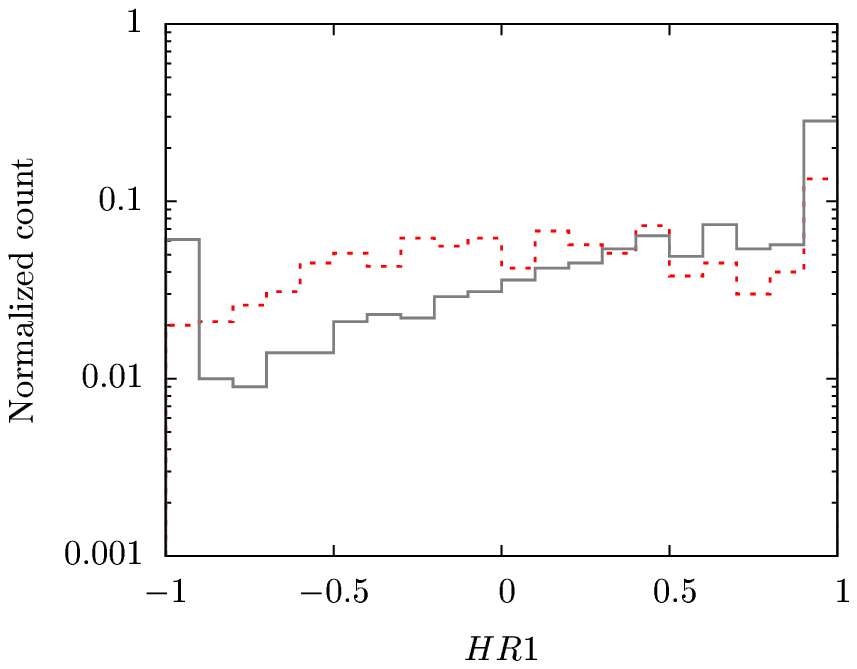}}
		\resizebox{40mm}{!}{\includegraphics[clip]{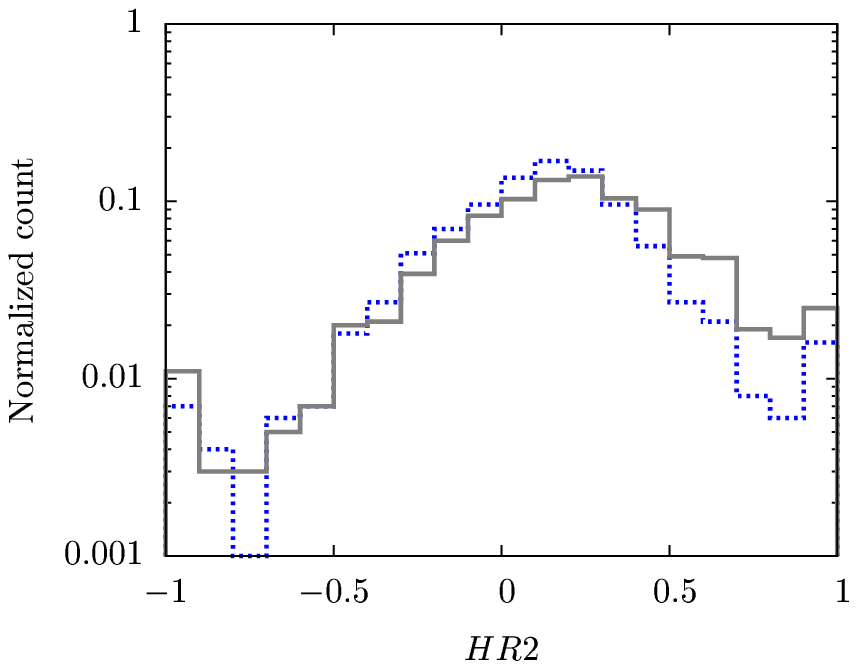}}
		\resizebox{40mm}{!}{\includegraphics[clip]{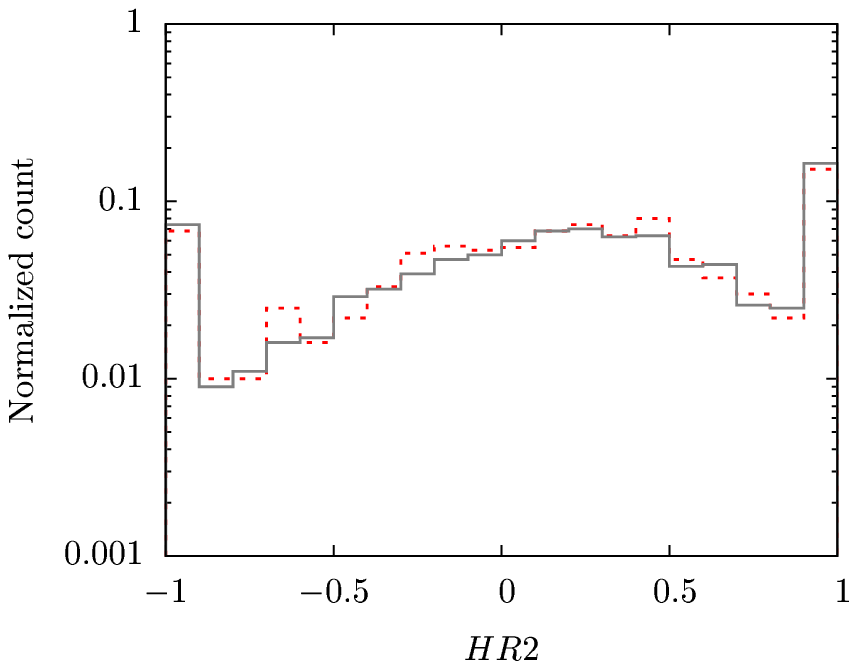}}
	\end{center}
		\caption{Histograms of hardness ration HR1 (top panel) and HR2 (bottom panel). 
					The number of sources are normalized. 
					The left panels are histograms of known AGNs (blue dotted line) and 
						unclassified sources (gray solid line) in the BSC. 
					The right panels are histograms of known AGNs (red dotted line) and 
						unclassified sources (gray solid line) in the FSC. 
  \label{HR-hist}}
\end{figure}

\begin{figure}
	\begin{center}
		\resizebox{40mm}{!}{\includegraphics[clip]{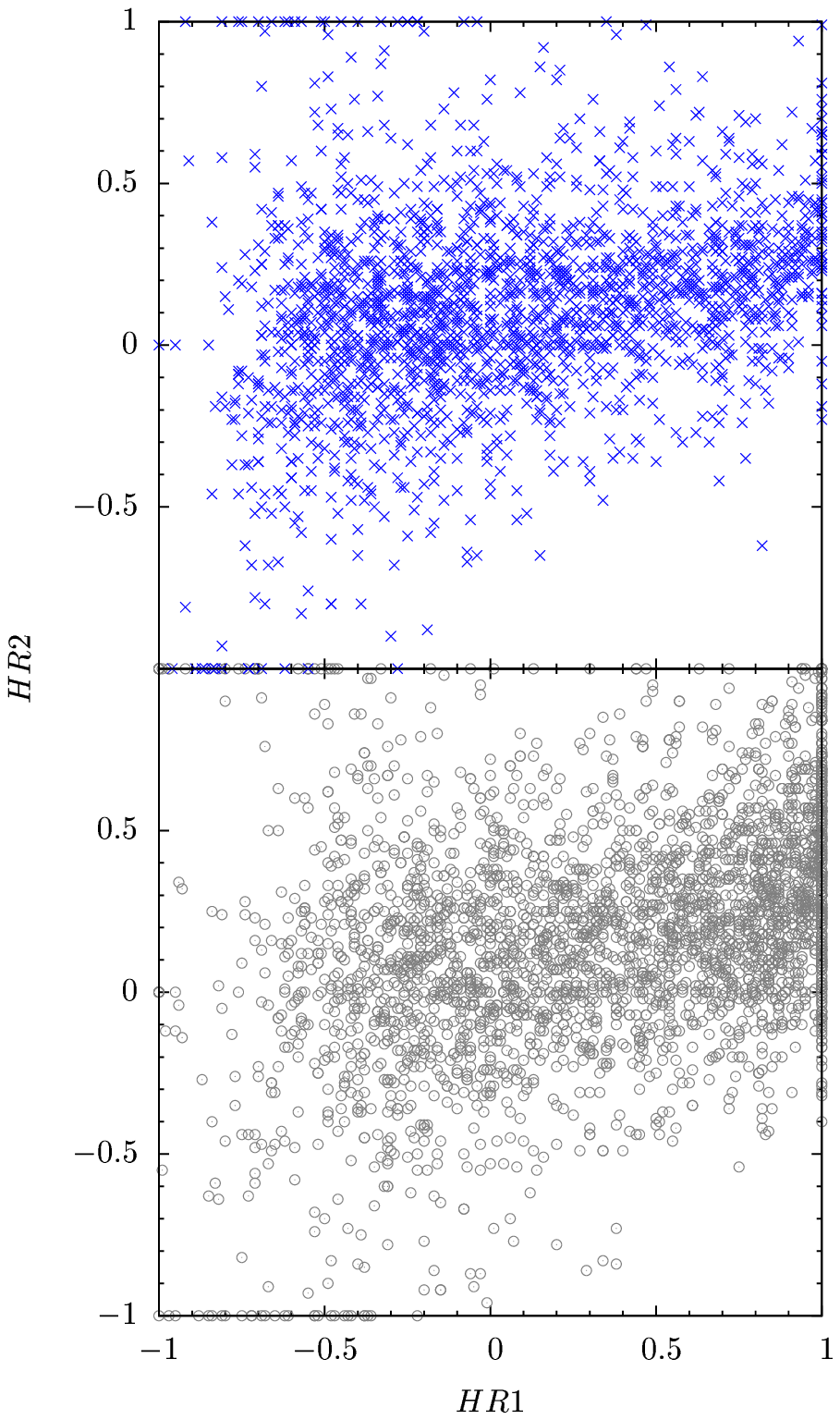}}
		\resizebox{40mm}{!}{\includegraphics[clip]{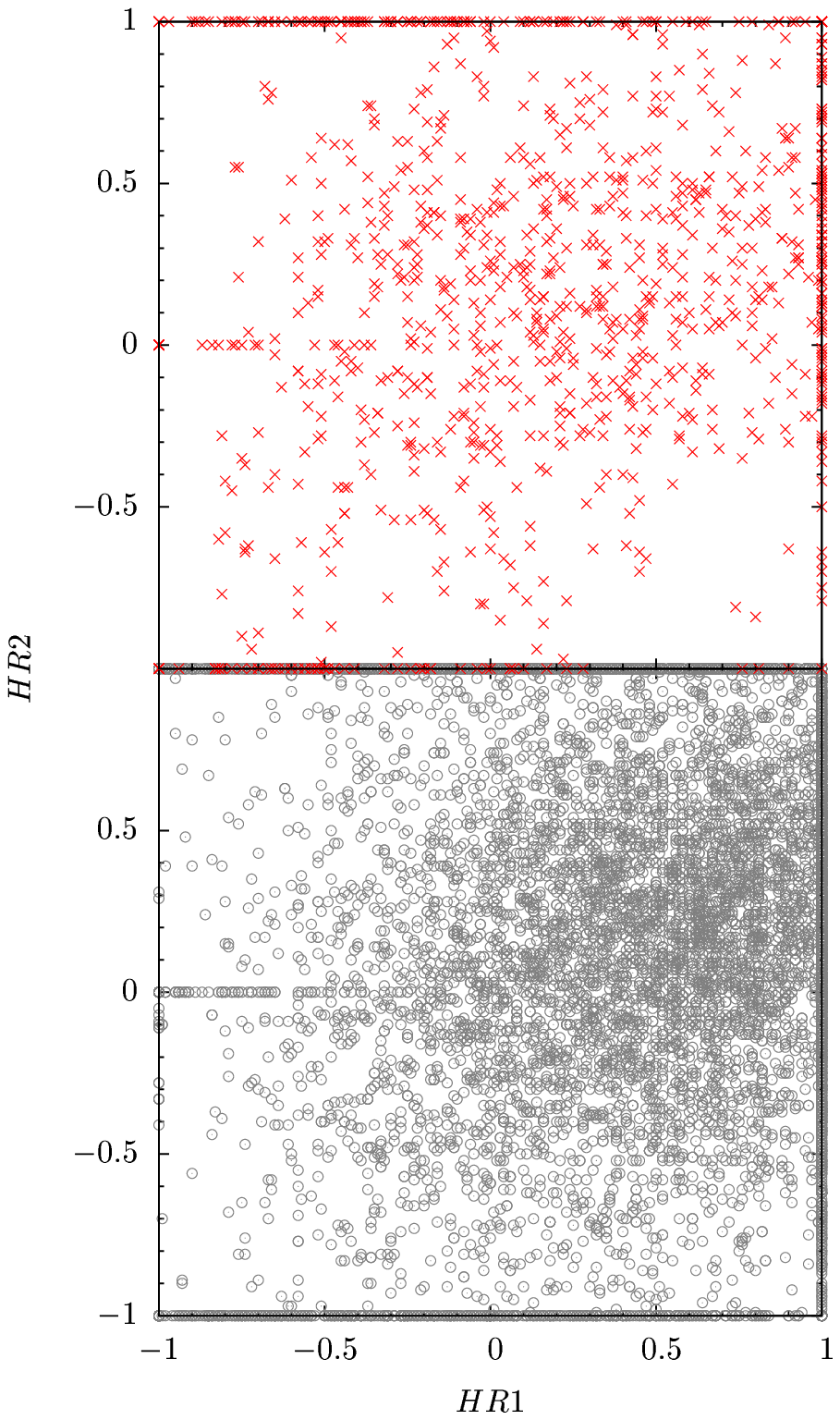}}
	\end{center}
		\caption{HR1 versus HR2 diagrams. 
					The symbols are same as in Figure \ref{Fx-K}.  \label{HR1-HR2}}
\end{figure}

Figure \ref{HR-hist} shows histograms of two hardness ratios (HR1 and HR2), 
where HR1 and HR2 are given using four energy bands, that is, 
A (Pulse Height Amplitude (PHA) channels 11--41), B (52--201), C (52--90), and D (91--201): 
HR1=(B-A)/(B+A) and HR2=(D-C)/(D+C), respectively. 
In the HR1 histogram for the BSC, both known AGNs and unclassified sources show
 a flat distribution in $-0.5 \la$HR1$\la 1.0$ and 
the number of sources decreases in HR1$\la-0.5$, 
although the number of unclassified sources is relatively larger in HR1$\ga 0.8$. 
\citet{Voges1999-AA} presented that X-ray counterparts for AGNs in the catalogue of \citet{Veron1998-BOOK} 
show a flat distribution in the range between $-0.5$ to $+1.0$. 
Hence, the HR1 property of these candidates is consistent with \citet{Voges1999-AA}. 
On the other hand, although known AGNs in the FSC have a similar shape in \citet{Voges1999-AA},
 the number of unclassified sources in the FSC increase towards larger HR1 values
 (i.e., they are relatively harder as compared to known AGNs). 
Therefore, some candidates in the FSC might be contaminated
 by galaxies because Abell clusters of galaxies (ACO) objects
 tend to have harder than stars and AGNs in their diagrams \citep{Voges1999-AA}. 
In the HR2 histograms, they have peaks at HR2$\sim0.2$ and
 this is also consistent with the HR2 histograms of AGNs in \citet{Voges1999-AA}. 
It should be noted that in the range over HR2$\sim0.2$ 
there are relatively large number of unclassified candidates in the BSC than known AGNs 
(i.e., there are harder samples) 
though the distributions for known/unknown sources in the FSC are alike. 

In Figure \ref{HR1-HR2}, HR1 versus HR2 diagrams are shown. 
Both known AGNs and unclassified sources in the BSC have similar distribution each other, 
although there are a relatively large number of unclassified sources in HR1$\ga0.8$ as pointed out the above. 
Most of them are located in HR1$>-0.6$ and $-0.5<$HR2$<0.5$ space and 
show a flat distribution with respect to HR1. 
\citet{Voges1999-AA} compared distributions among three types of objects 
(namely, TYCHO stars, ACO objects, and AGNs) and 
presented that AGNs are found mostly in the central part with HR1$>-0.5$ and $-0.5<$HR2$<	0.5$. 
The distribution of our samples are very similar to that of \citet{Voges1999-AA}. 
Therefore, our candidates in the BSC is consistent with the property of AGNs in \citet{Voges1999-AA}.
On the other hand, candidates in the FSC tend to have a dispersion distribution compared to candidates in the BSC. 
Although errors of hardness ratios affect the distribution, 
there may be many spurious AGNs in the candidates detected in the FSC.

%----------------------------
\subsection{Properties of Counterparts}

\begin{table}
\begin{center}
\caption{The number of counterparts extracted from each catalogue. 
			The numbers in parentheses are unclassified sources. 
\label{Counterparts-list}}
\begin{tabular}{llrr}
\hline \hline
catalogue & positional & number &  \\
 & criterion & BSC & FSC \\ \hline
SDSS   & $\leq 1''$ & 1,857 (624)   & 2,028 (1,301)  \\
DENIS  & $\leq 1''$ & 1,949 (1,443) & 3,902 (3,722) \\
FIRST  & $\leq 2''$ &   592 (106)   & 381 (159) \\
NVSS   & $\leq 2''$ &   534 (190)   & 383 (244) \\ \hline
\end{tabular}
\end{center}
\end{table}

AGNs are observable in multiwavelength. 
Accordingly, counterparts at other wavelengths provide us with information about their properties. 
To find multiwavelength counterparts of AGN candidates, 
the candidates were positionally cross-identified with the following catalogues: 
Sloan Digital Sky Survey (SDSS) Photometric Catalog, Release 7 \citep{Adelman2008-ApJS,Adelman2009-catalogue}, 
3rd release of the Deep Near Infrared Survey of the southern sky (DENIS) database (DENIS consortium, 2005),
 Faint Images of the Radio Sky at Twenty-centimeters (FIRST) Survey Catalog, Version 03Apr11 \citep{White1997-ApJ}, and 
1.4GHz National Radio Astronomy Observatory (NRAO) Very Large Array (VLA) Sky Survey \citep[NVSS;][]{Condon1998-AJ}. 
Table \ref{Counterparts-list} summarizes the number of counterparts in each catalogue.

\subsubsection{Optical Properties}

\begin{figure}
	\begin{center}
		\resizebox{40mm}{!}{\includegraphics[clip]{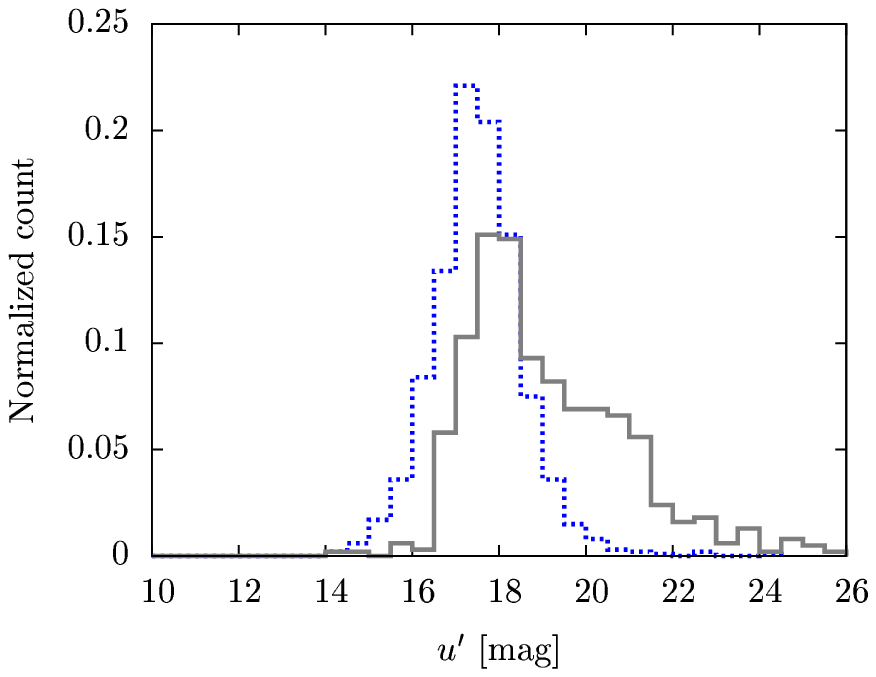}}
		\resizebox{40mm}{!}{\includegraphics[clip]{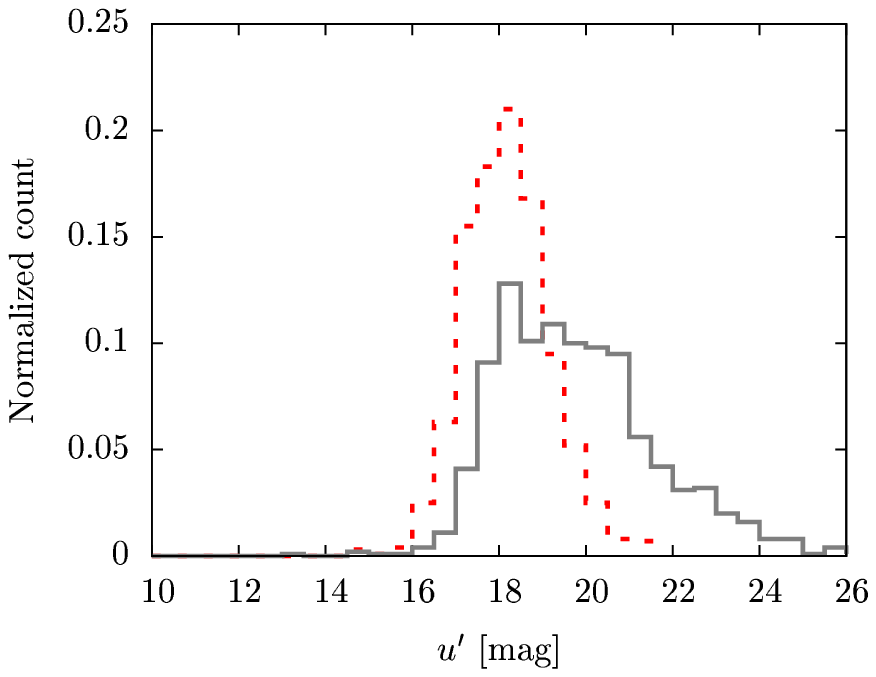}}
	\end{center}
		\caption{Histograms of $u'$ magnitudes. 
					The number of sources are normalized. 
					The symbols are the same as in Figure \ref{HR-hist}. 
  \label{u-hist}}
\end{figure}

\begin{figure}
	\begin{center}
		\resizebox{40mm}{!}{\includegraphics[clip]{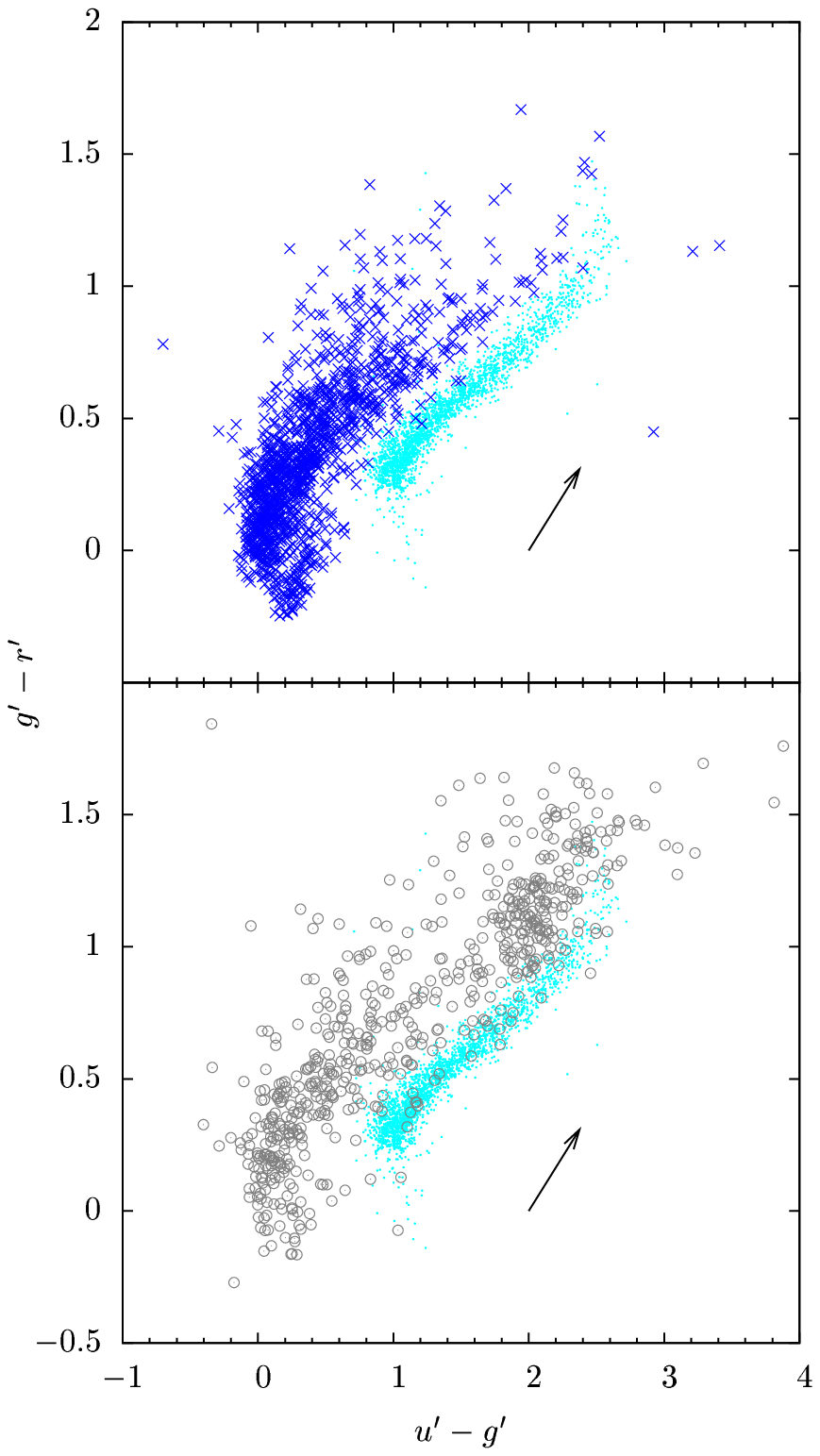}}
		\resizebox{40mm}{!}{\includegraphics[clip]{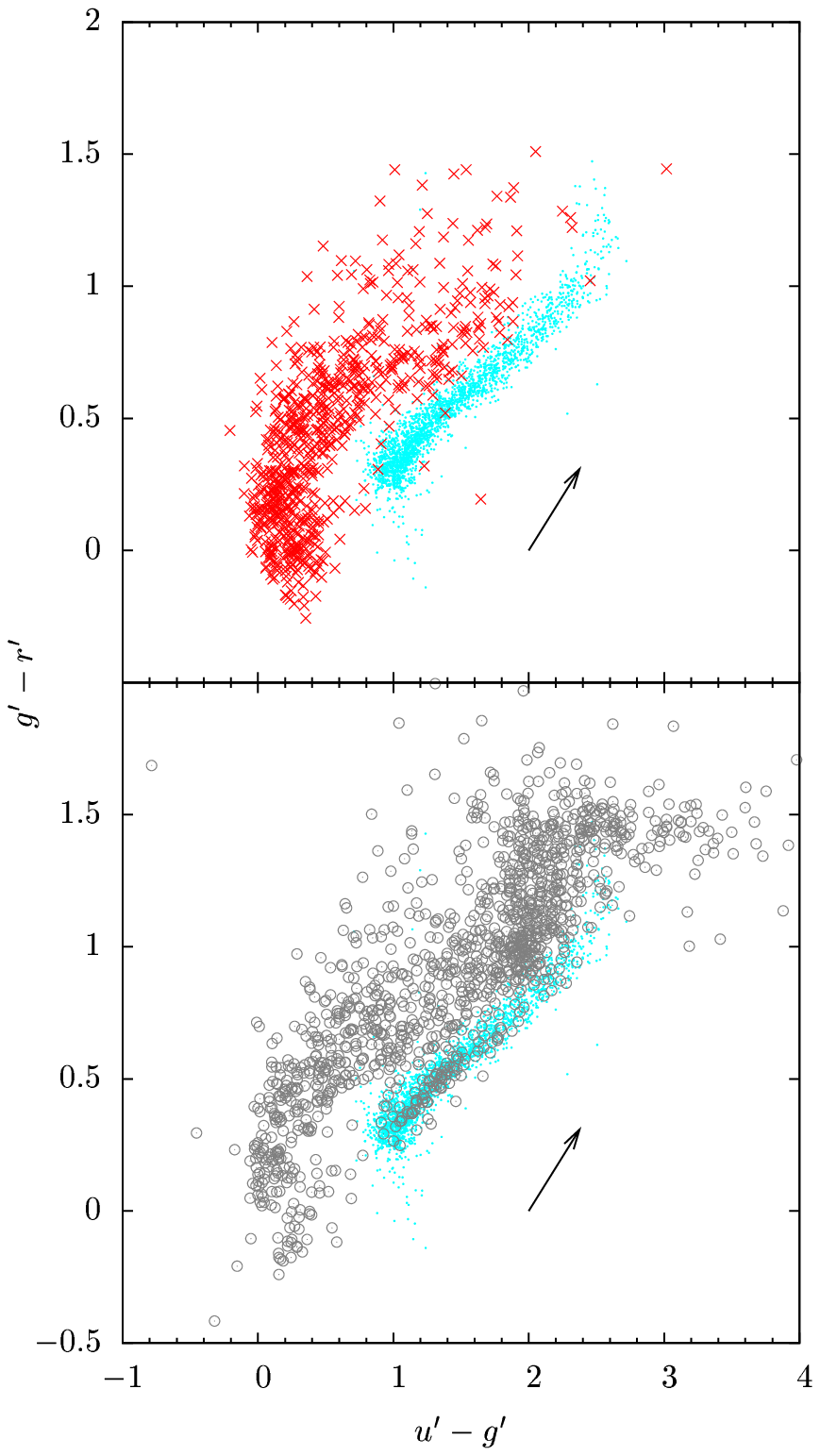}}
	\end{center}
		\caption{Optical CCD using SDSS photometry. 
					The light-blue dots represent normal stars. 
					The other symbols are the same as in Figure \ref{Fx-K}. 
					The reddening vector is also shown in the diagrams. 
  \label{ug-gr-CCD}}
\end{figure}

Of 1,857 (2,028) SDSS counterparts in the BSC (FSC), 624 (1,301) are unclassified sources. 
Here, we investigate photometric properties using SDSS photometry. 

Figure \ref{u-hist} shows histograms of $u'$ magnitudes in the SDSS catalogue. 
Unclassified sources are relatively fainter than known AGNs. 
This is because the search in the near-infrared/X-ray, in which light suffers smaller extinction than in the optical,
 enables us to detect faint sources in the optical wavelength. 

In Figure \ref{ug-gr-CCD}, we plotted candidates on a $(u'-g')$-$(g'-r')$ CCD. 
There are several studies investigating stellar locus in a CCD using SDSS photometry 
\citep[e.g., ][]{Fan1999-AJ,Finlator2000-AJ,Sesar2006-AJ}. 
Their stellar loci are consistent with each other. 
\citet{Anderson2003-AJ,Anderson2007-AJ} also investigated distributions of normal stars,
 AGNs, and quasars in the $(u'-g')$-$(g'-r')$ CCD, and
 demonstrated that loci of quasars/AGNs are separated from the stellar locus. 
The locus of sample stars, taken from the SDSS catalogue, is also shown in Figure \ref{ug-gr-CCD}. 
The reddening vector is based on \citet{Fukugita2004-AJ}. 
Almost all of the known AGNs in both catalogues are clearly separated from the stellar locus. 
Although most unclassified sources are not located at the stellar locus, 
some sources are distributed around the stellar locus. 
It is noticeable for unclassified sources in the FSC.  
These sources might not be AGNs but normal stars. 

We can notice that some unclassified sources are located in $(u'-g')>1.5$ (especially, around $(u'-g')\sim 2.0$), 
where few known AGNs are distributed. 
Most of these sources are still differentiated from the stellar locus. 
These might be highly reddened AGNs or other types of AGNs that are not discovered in SDSS surveys. 
We note that most of the unclassified sources with $(u'-g')<1.5$ are probably newly discovered AGNs. 

\if2-----------------------
\vspace{5mm}
This is seen from histograms of SDSS magnitudes. 
Figure \ref{u-hist} shows histograms of $u'$ and $z'$ magnitudes. 
Although there is no difference between known and unknown sources in $z'$ histograms, 
a clear difference is seen in $u'$ histograms. 
This is probably due to a reddening. 
Therefore, the fact that unknown sources have relatively redder colours than known AGNs 
is consistent with 
\fi

%------------------
\subsubsection{Near-Infrared Variability}

\begin{figure}
	\begin{center}
		\resizebox{40mm}{!}{\includegraphics[clip]{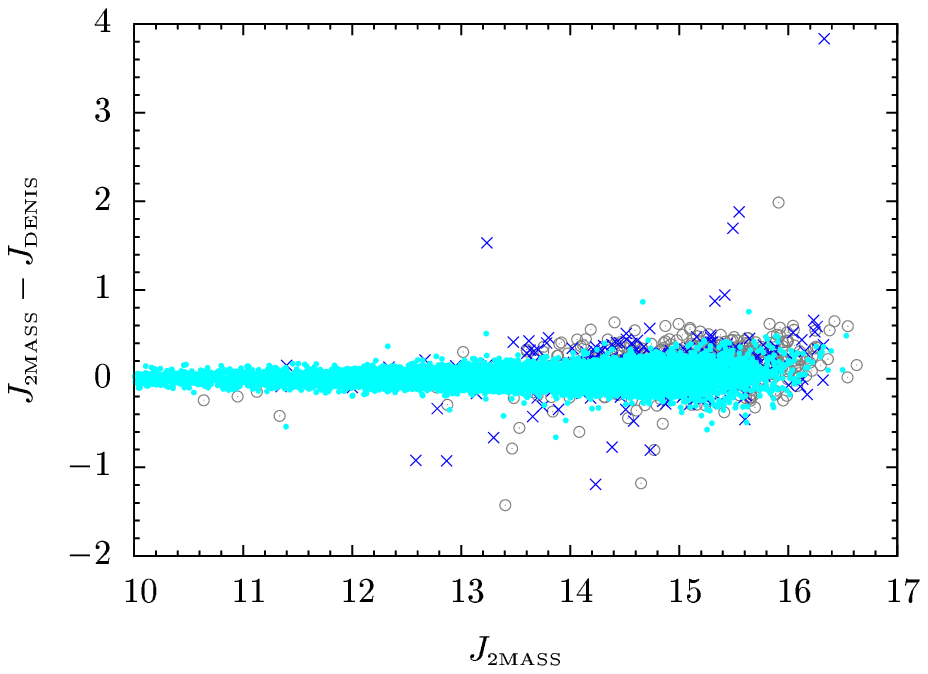}}
		\resizebox{40mm}{!}{\includegraphics[clip]{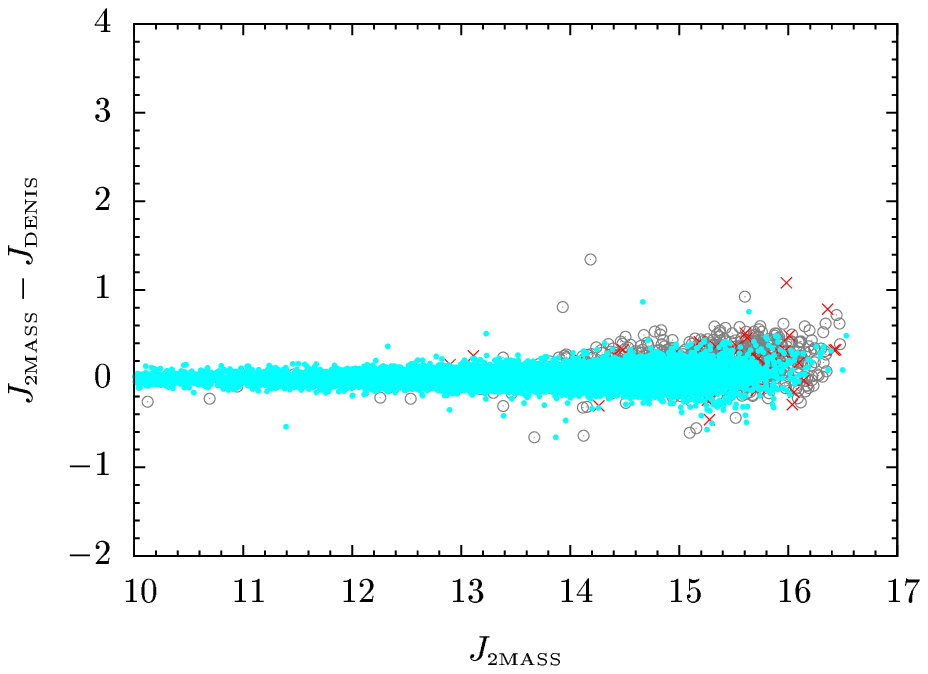}}
		\resizebox{40mm}{!}{\includegraphics[clip]{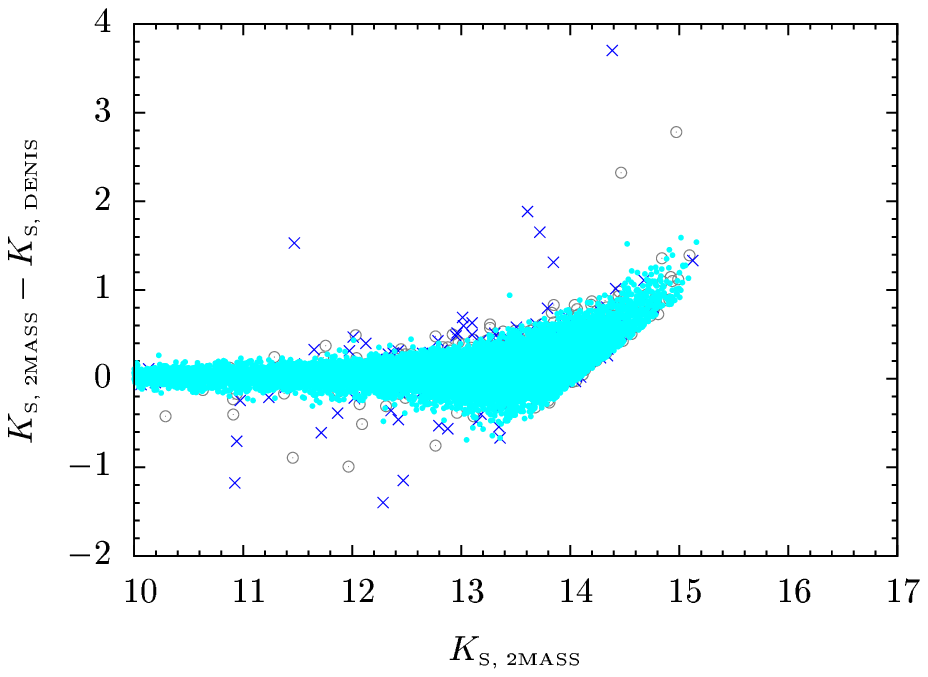}}
		\resizebox{40mm}{!}{\includegraphics[clip]{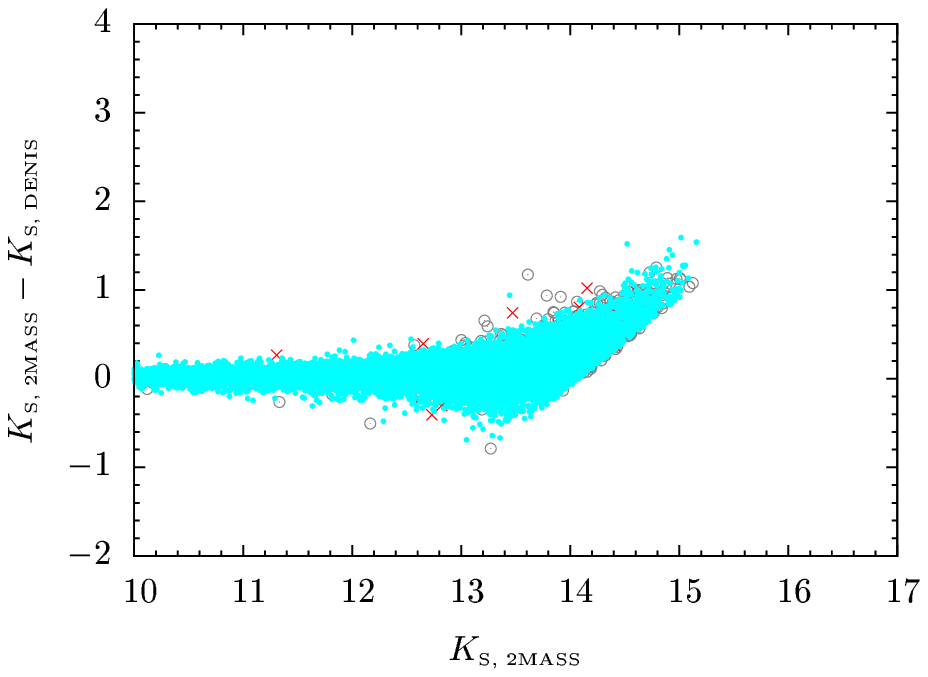}}
	\end{center}
		\caption{$J_\textnormal{\tiny 2MASS}$ versus $(J_\textnormal{\tiny 2MASS}-J_\textnormal{\tiny DENIS})$ and 
					$K_\textnormal{\tiny S, 2MASS}$ versus $(K_\textnormal{\tiny S, 2MASS}-K_\textnormal{\tiny S, DENIS})$ diagrams. 
					The symbols are the same as in Figure \ref{ug-gr-CCD}. 
 \label{mag-deltamag}}
\end{figure}

\begin{table}
\caption{The number of sources with near-infrared variability. 
			The values in parentheses represent percentages of total number of sources.  
\label{Variability-list}}
\begin{center}
\begin{tabular}{lrrrr}
\hline \hline
criterion              & \multicolumn{2}{c}{BSC} & \multicolumn{2}{c}{FSC}  \\
                       & known & unclassified & known & unclassified  \\ \hline
$>3\sigma_{\Delta J}$  & 98 (19) & 211 (15) & 25 (14) & 256 (7) \\
$>3\sigma_{\Delta K_\textnormal{\tiny S}}$ & 24 (5) & 73 (5) & 10 (6) & 148 (4) \\
$>5\sigma_{\Delta J}$  & 36 (7) & 62 (4) & 7 (4) & 77 (2) \\
$>5\sigma_{\Delta K_\textnormal{\tiny S}}$ & 11 (2) & 8 (1) & 1 (1) & 16 (0) \\ \hline
\end{tabular}
\end{center}
\end{table}

AGNs are known to have variability across wide wavelength ranges from radio to X-ray or $\gamma$-ray 
\citep{Krolik1991-ApJ,Edelson1996-ApJ,Giveon1999-MNRAS}. 
Here, we investigate variability of AGN candidates in the near-infrared wavelength, 
by comparing magnitudes between 2MASS and DENIS. 

Figure \ref{mag-deltamag} shows $J$-$\Delta J$ and $K_\textnormal{\tiny S}$-$\Delta K_\textnormal{\tiny S}$ diagrams, 
where $\Delta J =J_\textnormal{\tiny 2MASS}-J_\textnormal{\tiny DENIS}$ and
 $\Delta K_\textnormal{\tiny S} =K_\textnormal{\tiny S, 2MASS}-K_\textnormal{\tiny S, DENIS}$, respectively. 
We also plotted photometric differences of normal stars for referring to typical difference
 between 2MASS and DENIS photometric systems. 
Sample stars are taken from $l=260^\circ, b=+3^\circ$ area,
 whose near-infrared colours are consistent with those of normal stars in \citet{Bessell1988-PASP}. 
We note that the sample stars have photometric quality flags in both 2MASS and DENIS catalogues 
superior to B ($S/N>7$) and 90, respectively. 
The DENIS photometry is transformed into the 2MASS photometry 
on the basis of \citet{Carpenter2001-AJ}. 
Standard deviations $\sigma_{\Delta J}$ and $\sigma_{\Delta K_\textnormal{\tiny S}}$
 for the sample stars are 0.08 and 0.20, respectively. 
As seen in the diagrams, there are several sources clearly showing variability. 
Table \ref{Variability-list} summarizes the number of candidates showing variability larger than $3\sigma$ and $5\sigma$. 
In our sample, percentages of candidates with variability in the FSC are relatively smaller than those in the BSC. 
Faint AGNs at the X-ray possibly show less variability. 
We note that, practically, there should be more sources with variability
 because we could compare only double-epoch magnitudes.

\subsubsection{Radio Properties}

\begin{figure}
	\begin{center}
		\resizebox{40mm}{!}{\includegraphics[clip]{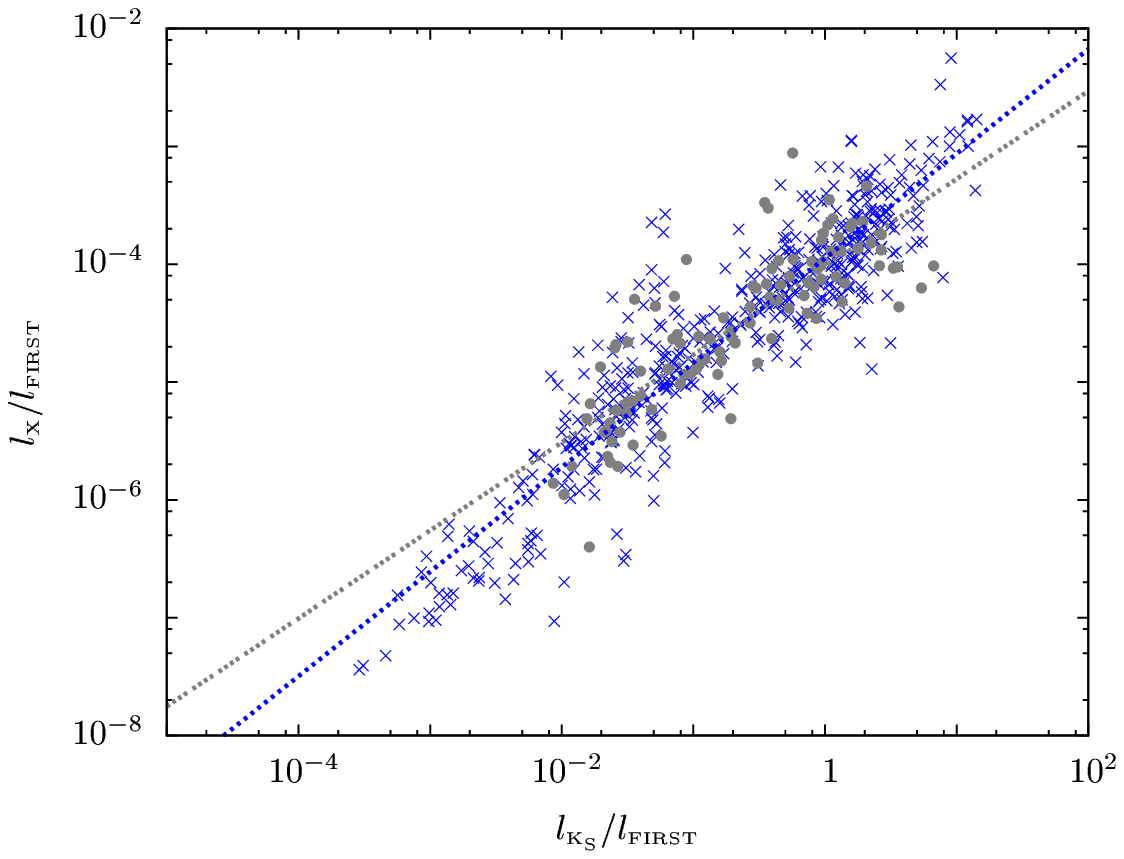}}
		\resizebox{40mm}{!}{\includegraphics[clip]{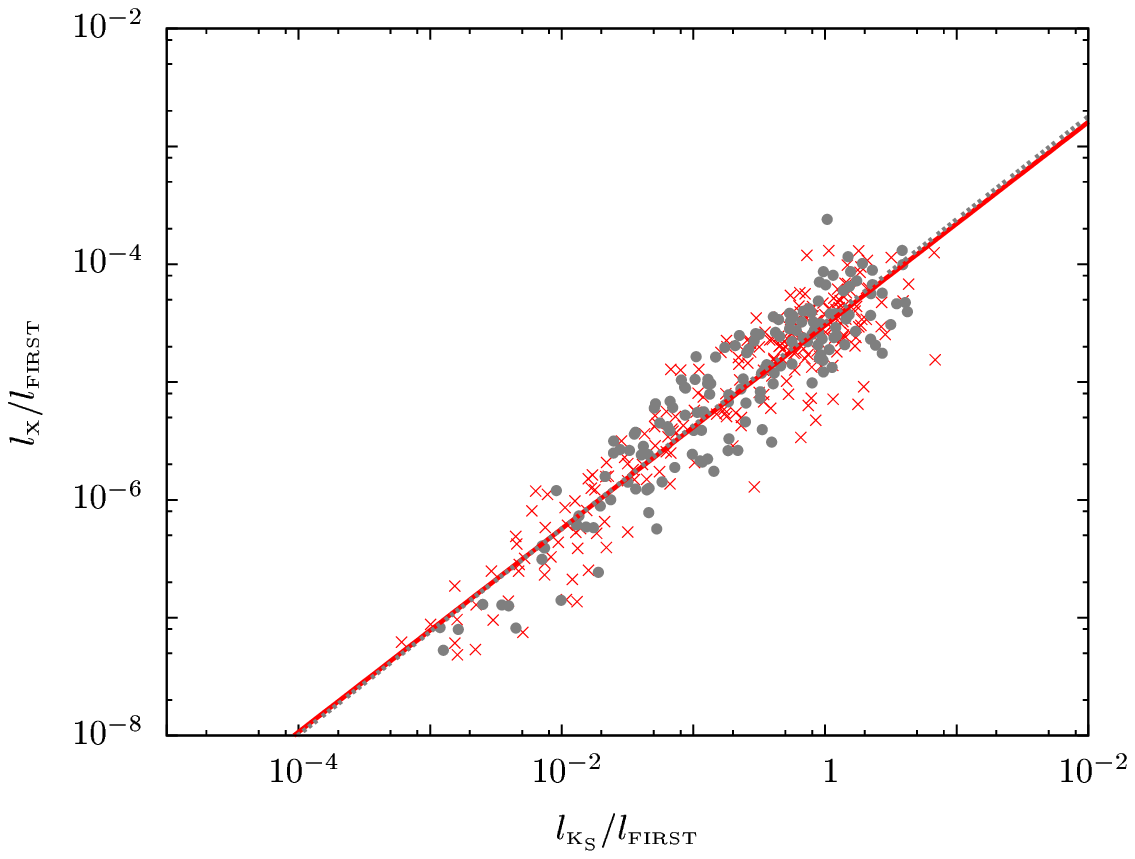}}
		\resizebox{40mm}{!}{\includegraphics[clip]{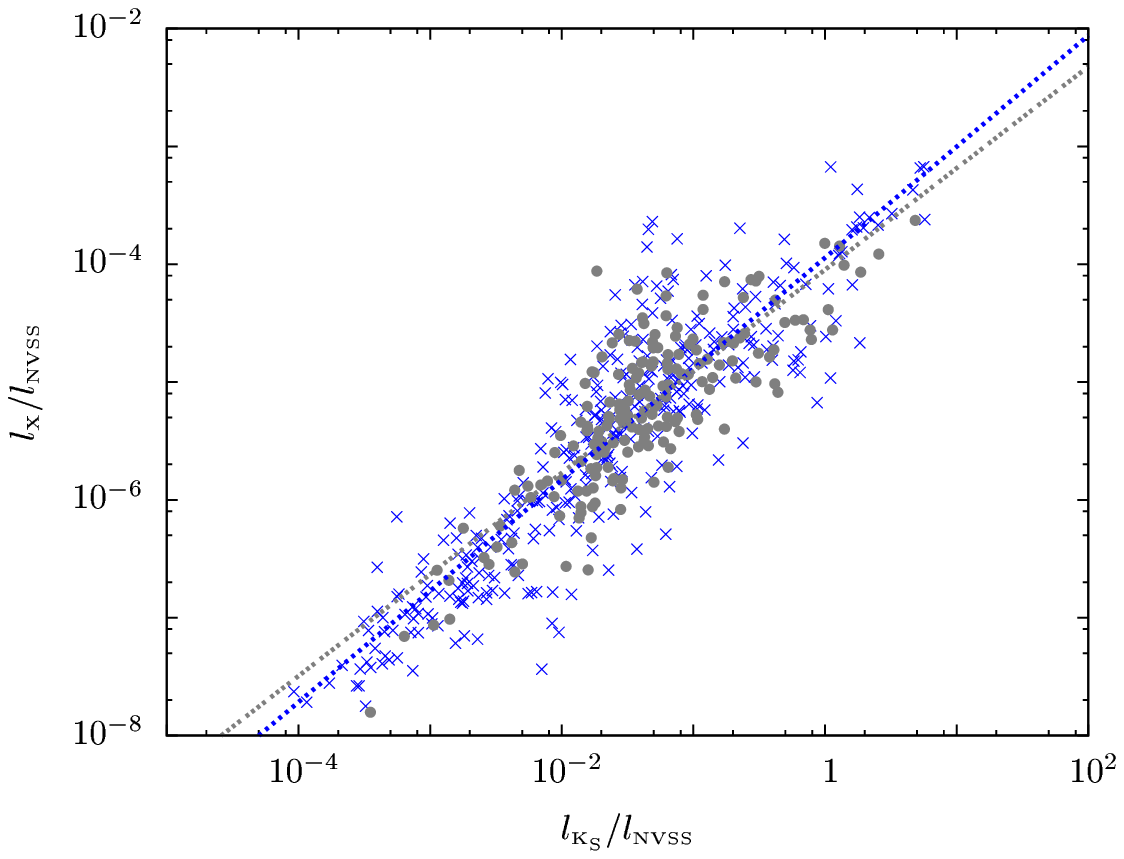}}
		\resizebox{40mm}{!}{\includegraphics[clip]{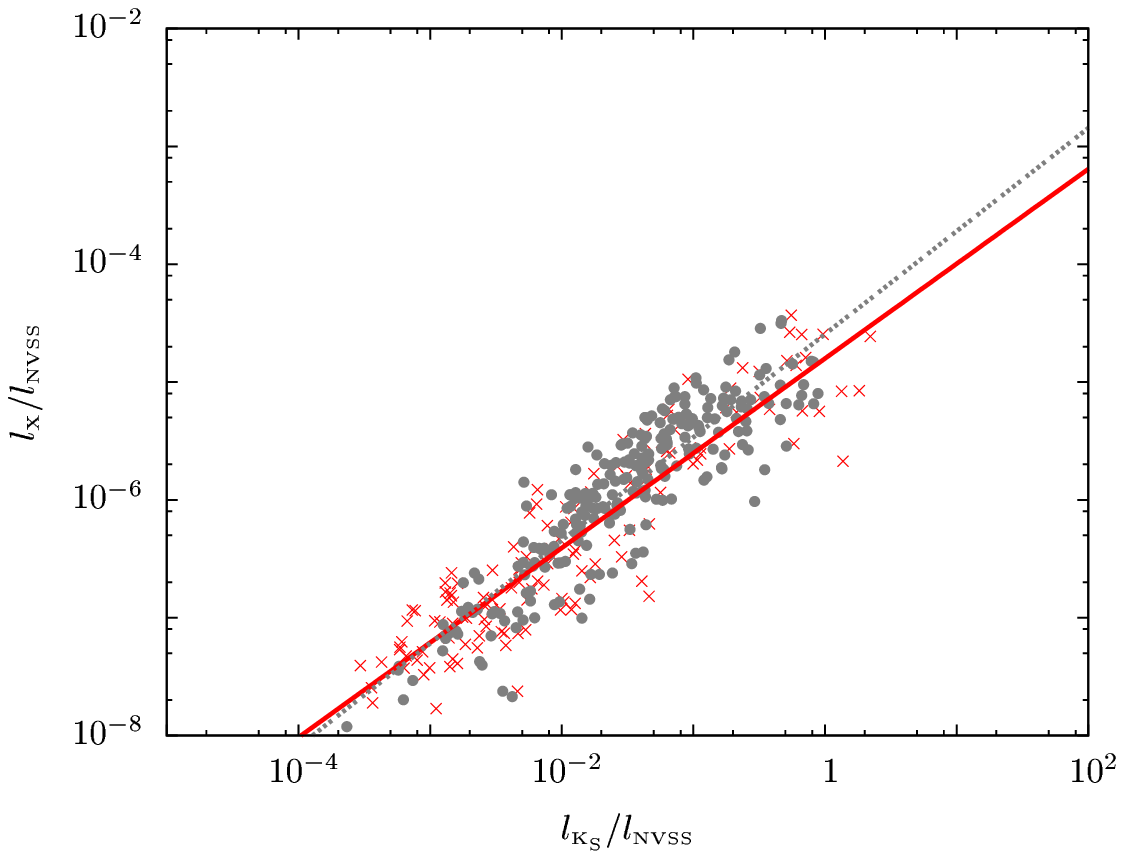}}
	\end{center}
		\caption{Luminosity ratios $F_\textnormal{\tiny K$_\textnormal{\tiny S}$}/F_\textnormal{\tiny radio}$ versus
				 $F_\textnormal{\tiny X}/F_\textnormal{\tiny radio}$ diagram. 
				 The upper panels are diagrams for FIRST counterparts, and the lower panels are those for NVSS counterparts. 
				 The lines are derived by the least square method. 
 \label{krad-xrad-CCD}}
\end{figure}

Figure \ref{krad-xrad-CCD} shows flux ratios
 $l_\textnormal{\tiny K$_\textnormal{\tiny S}$}/l_\textnormal{\tiny radio}$ versus
 $l_\textnormal{\tiny X}/l_\textnormal{\tiny radio}$ diagram. 
Significant differences between known AGNs and unclassified sources can not be clearly seen in each diagram. 
There appears to be correlations between $l_\textnormal{\tiny K$_\textnormal{\tiny S}$}$ and
 $l_\textnormal{\tiny radio}$ in the diagrams. 
We performed least square fittings for the distributions of both known AGNs and unclassified sources. 
The least square fitting lines are also shown in the diagrams. 
Still, there is no significant difference between known AGNs and unclassified sources in each diagram, 
although in the $l_\textnormal{\tiny K$_\textnormal{\tiny S}$}/l_\textnormal{\tiny FIRST}$--
$l_\textnormal{\tiny X}/l_\textnormal{\tiny FIRST}$ for the BSC there is small difference in slope. 
It is known that there is a correlation between luminosity ratios $l_\textnormal{o}/l_\textnormal{r}$ 
and $l_\textnormal{x}/l_\textnormal{r}$ \citep[e.g., ][]{Brinkmann1995-AAS,Brinkmann1997-AA,Brinkmann2000-AA}. 
Correlations in our sample may be a variant of such correlations. 

Because unclassified sources have similar properties with known AGNs in the diagrams, 
these properties also support that unclassified sources are AGNs.

%%%%%%%%%%%%%%%%%%%%%%%%%%%%%%%%%%%%%%%
\section{Summary and Conclusions}
We have cross-identified the 2MASS PSC with ROSAT catalogues, and 
have extracted AGN candidates on the basis of the near-infrared colour selection. 
Of 5,273 (10,701) AGN candidates in the BSC (FSC), 3,220 (9,693) are unclassified sources. 
We investigated their properties using near-infrared, X-ray, optical, and radio data. 
Overall, most unclassified sources in the BSC have similar properties with known AGNs and 
their properties are also consistent with previous studies, 
whereas some unclassified sources, especially those in the FSC, have properties different from known AGNs. 

It is highly probable that most unclassified sources are AGNs because of the following reasons. 
First, our candidates are X-ray sources. 
The fact that a object is a X-ray source supports the object is an AGN 
because it is believed that the vast majority of X-ray sources are AGNs. 
Second, our candidates satisfy the near-infrared colour selection criteria
 proposed by \citet{Kouzuma2010-AA}, that is, 
they have properties in near-infrared colours similar to those of AGNs. 
Third, some candidates have been already known AGNs and
 not only unclassified sources show properties similar to those of known AGNs
 but also their properties in multiwavelength are consistent with previous studies. 
However, it should be noted that there may be 
relatively large number of contamination in unclassified FSC sources. 
To confirm that they are real AGNs, spectroscopic observations are required. 

Similar works using a cross-identification with a ROSAT catalogue were performed 
by \citet{Boller1992-AA2,Boller1998-AAS}. 
\citet{Boller1992-AA2} extracted a sample of 14,708 extragalactic IRAS sources 
on the basis of a supervised selection 
whose selection quality was assessed by \citet{Boller1992-AA}, and 
they cross-identified between the sample and the first processing of the ROSAT all-sky survey data. 
The cross-identified 244 IRAS galaxies comprised infrared luminous nearby galaxies, 
spirals and ellipticals, Seyfert galaxies, and QSOs. 
They investigated properties of the cross-identified sources using X-ray, far-infrared, and optical fluxes, and 
found some correlations between them. 
The work in \citet{Boller1998-AAS} is also similar to that in \citet{Boller1992-AA2}, 
which cross-indentified the 14,315 IRAS galaxies with the second processing of the ROSAT all-sky survey. 
These studies are similar to our study in that a ROSAT catalogue is cross-identified with an infrared catalogue. 
However, whereas these studies investigated extragalactic sources 
extracted by a mid- and far-infrared colour selection, 
we investigated AGN candidates extracted by the near-infrared colour selection. 
The number of sample is also totally different. 
In addition, we discussed not only properties in infrared and X-ray 
but also near-infrared variability and radio properties, 
which are not discussed in \citet{Boller1992-AA2,Boller1998-AAS}. 

%different in sample selection and 

This paper extracted several thousands of AGN candidates across the entire sky.
The use of only near-infrared colours enables us to search for AGN candidates across the entire sky. 
When using this technique on other deep near-infrared surveys such as DENIS, UKIDSS, and future surveys, 
many AGNs, which have not been detected in the optical wavelength, can be extracted.

%These large number of samples may be useful for 

\section*{Acknowledgement}
This publication makes use of data products from the Two Micron All Sky Survey, 
which is a joint project of the University of Massachusetts and 
the Infrared Processing and Analysis Center/California Institute of Technology, 
funded by the National Aeronautics and Space Administration and the National Science Foundation.
TOPCAT (http://www.starlink.ac.uk/topcat/) helps us investigate
 the photometric properties of the AGN candidates. 
We thank the referee, Dr. Thomas Boller, for his suggestions.

\bibliographystyle{mn2e}

\bsp
\label{lastpage}
\end{document}